\documentclass[aps, twocolumn, prl, showpacs, amsmath, superscriptaddress]{revtex4}
\usepackage{graphicx}
\usepackage{dcolumn}
\usepackage{bm}
\usepackage{amsmath}
\usepackage{subfigure}
\usepackage{amsfonts}
\usepackage{xcolor}
\usepackage{appendix}
\usepackage{multirow}

\newcommand{\ket}[1]{$|#1\rangle$}

\begin{document}

\title{The classical nature of nuclear spin noise near clock transitions of Bi donors in silicon}
\author{Wen-Long Ma$^{\ddag,}$}
\affiliation{State Key Laboratory of Superlattices and Microstructures, Institute of Semiconductors, Chinese Academy of Sciences, Beijing 100083, China}
\affiliation{Department of Physics, The Chinese University of Hong Kong, Shatin, N. T., Hong Kong, China}
\affiliation{Synergetic Innovation Center of Quantum Information and Quantum Physics, University of Science and Technology of China, Hefei, Anhui 230026, China}
\author{Gary Wolfowicz$^{\ddag,}$}
\affiliation{London Centre for Nanotechnology, University College London, London WC1H 0AH, UK}
\affiliation{Department of Materials, Oxford University, Oxford OX1 3PH, UK}
\author{Shu-Shen Li}
\affiliation{State Key Laboratory of Superlattices and Microstructures, Institute of Semiconductors, Chinese Academy of Sciences, Beijing 100083, China}
\affiliation{Synergetic Innovation Center of Quantum Information and Quantum Physics, University of Science and Technology of China, Hefei, Anhui 230026, China}
\author{John J.L. Morton}
\email{jjl.morton@ucl.ac.uk}
\affiliation{London Centre for Nanotechnology, University College London, London WC1H 0AH, UK}
\affiliation{Department of Electronic and Electrical Engineering, University College London, London WC1E 7JE, UK}
\author{Ren-Bao Liu}
\email{rbliu@phys.cuhk.edu.hk}
\affiliation{Department of Physics, The Chinese University of Hong Kong, Shatin, N. T., Hong Kong, China}
\affiliation{Center for Quantum Coherence, The Chinese University of Hong Kong, Shatin, N.T., Hong Kong, China}
\affiliation{Institute of Theoretical Physics, The Chinese University of Hong Kong, Shatin, N.T., Hong Kong, China}


\date{\today }

\begin{abstract}
Whether a quantum bath can be approximated as classical noise is a fundamental issue in central spin decoherence and also of practical importance in designing noise-resilient quantum control. Spin qubits based on bismuth donors in silicon have tunable interactions with nuclear spin baths and are first-order insensitive to magnetic noise at so-called clock-transitions (CTs). This system is therefore ideal for studying the quantum/classical nature of nuclear spin baths since the qubit-bath interaction strength determines the back-action on the baths and hence the adequacy of a classical noise model. We develop a Gaussian noise model with noise correlations determined by quantum calculations and compare the classical noise approximation to the full quantum bath theory.
 We experimentally test our model through dynamical decoupling sequence of up to 128 pulses, finding good agreement with simulations and measuring electron spin coherence times approaching one second --- notably using natural silicon. Our theoretical and experimental study demonstrates that the noise from a nuclear spin bath is analogous to classical Gaussian noise if the back-action of the qubit on the bath is small compared to the internal bath dynamics, as is the case close to CTs. However, far from the CTs, the back-action of the central spin on the bath is such that the quantum model is required to accurately model spin decoherence.

\end{abstract}
\maketitle

\textit{Introduction}. Central spin decoherence due to coupling to the environment is not only a central issue in understanding quantum-to-classical transitions~\cite{De-1,De-2}, but also one of the key challenges in the realization of quantum computation~\cite{De-3}. There are two distinct models to describe the decoherence processes in such cases: in the semiclassical model, the central spin accumulates random phases due to thermal or quantum fluctuations of the environment~\cite{cla-1, cla-2}, while in the quantum model, the coupling between the central spin and the environment produces entanglement and results in leakage of the which-way information from the central spin to the environment \cite{quan-CE, quan-pair, quan-CCE, quan-adjoint}. The fundamental difference between these two models lies in the fact that the classical noise is independent of the central spin state while the quantum noise is governed by the back-action from the central spin \cite{NV-1,NV-2}.

The classical noise model of quantum baths is a useful approximation in designing noise-resilient quantum control \cite{Corre}, which would otherwise require a large amount of numerical simulations of many-body dynamics of quantum baths. Dynamical decoupling has been employed to extract the noise spectra of baths \cite{spec-1,spec-2,spec-3,spec-4}, which are in turn used to design optimal quantum control for protecting quantum coherence and quantum gates. The viability of such methods critically depends on whether the noise picture is valid or not. Therefore, examining the conditions for the validity of classical noise model is not only of fundamental interest but also highly desirable for accurate quantum control under realistic conditions.

Bismuth donors in silicon (Si:Bi) have recently attracted much attention in spin-based quantum computation due to a number of favourable properties~\cite{George10,Morley10,Bi-2}. These include  long electron spin coherence times of up to three seconds~\cite{Bi-1} observed for Bi donors (in isotopically-enriched silicon-28) tuned to so-called clock transitions (CTs)  --- also known as optimal working points (OWPs)~\cite{Bi-4} or zero first-order Zeeman (ZEFOZ) transitions) --- whose frequency is insensitive, to first order, to magnetic field fluctuations. The coherence times of donor electron spins  in natural silicon is typically limited to a few hundred microseconds by the 5\% naturally abundant $^{29}$Si nuclear spins. At CTs, the effect of the $^{29}$Si on the electron spin coherence is strongly suppressed (though not completely removed), leading to coherence times of up to 100 ms. Previous studies have focused on quantum approaches to model electron spin decoherence from $^{29}$Si nuclear spin baths~\cite{Bi-5, Bi-6}, however, performing such calculations near the CTs can be challenging due to the strongly correlated nuclear spin baths~\cite{Bi-5}.

In this Letter, we explore the applicability of a classical Gaussian stochastic noise description~\cite{Si-1,Gauss} of the nuclear spin bath around Bi donors in natural silicon, especially near the Si:Bi CTs. Such classical Gaussian noise is fully characterized by the two-point correlation functions or the noise spectrum~\cite{Gauss}. We demonstrate the validity of  such a semiclassical model by comparisons with exact results from the quantum model and with experimental measurements.

\textit{System and Hamiltonian}. For the Si:Bi system interacting with a $^{29}$Si nuclear spin bath ($I_{i}=1/2$ and natural abundance 4.7$\%$ throughout the host lattice), the system Hamiltonian is divided into three parts~\cite{Bi-5,Bi-6}:
\begin{eqnarray}\label{H}
 H=H_{\text{cs}}+H_{\text{int}}+H_{\text{bath}},
 \label{}
\end{eqnarray}
with
\begin{subequations}\label{eq:1}
\begin{align}
& H_{\text{cs}}=\omega_{e}S^{z}-\omega_{n}^{\rm Bi}{I}_{0}^{z}+A_{0}\mathbf{S}\cdot\mathbf{I}_{0},          \label{A} \\
& H_{\text{int}}=S^{z}\sum_{i}A_{i}I_{i}^{z}, \label{B}  \\
& H_{\text{bath}}=-\omega_{n}^{\rm Si}\sum_{i}{I_{i}^{z}}+\sum_{i<j}\mathbf{I}_{i}\cdot\mathbb{D}_{ij}\cdot\mathbf{I}_{j},           \label{C}
\end{align}
\end{subequations}
where $\mathbf{S}$ is the Bi donor electron spin operator, $\mathbf{I}_{0} (\mathbf{I}_{i})$ is the $^{209}$Bi ($^{29}$Si) nuclear spin operator, $\omega_{\text{e}}$, $\omega_{\text{n}}^{\rm Bi}$ and $\omega_{\text{n}}^{\rm Si}$ are correspondingly the Larmor frequencies of the donor electron spin, $^{209}$Bi and $^{29}$Si nuclear spins (which are related to their gyromagnetic ratios $\gamma$ by $\omega_{\alpha}= \gamma_{\alpha} B$),
$A_{0}(A_{i})$ is the coupling strength between the donor electron spin and the $^{209}$Bi($^{29}$Si) nuclear spins, $\mathbb{D}_{ij}$ is the nuclear-nuclear interaction tensor, and $B$ is the magnetic field applied along the $z$ axis. Here we have neglected the non-secular terms in $H_{\text{int}}$ which induce the central spin relaxation, because the qubit energy splitting is much larger than the qubit-bath coupling (see online Supplementary Materials for detailed discussions on the effects of the non-secular terms on the pure-dephasing model)\cite{SI}.

\textit{Semiclassical model for quantum decoherence}.
The combined Bi donor electron-nuclear spin system ($S=1/2$, and $I_{0}=9/2$) with the Hamiltonian $H_{\text{cs}}$ has 20 eigenstates with eigenenergies dependent on the magnetic field \cite{SI}. By focusing on a pair of the eigenstates, \ket{+} and \ket{-}, of the central spin, we can recast the system Hamiltonian in Eq.~(\ref{H}) as a function of these two central spin states:
\begin{eqnarray}\label{}
 H^{(\pm)}=\pm\frac{P_{+}-P_{-}}{2}\beta^{z}+\frac{P_{+}+P_{-}}{2}\beta^{z}+H_{\text{bath}},
 \label{Hpm}
\end{eqnarray}
where $\beta^{z}=\sum_{i}\beta^{z}_{i}=\sum_{i}A_{i}I_{i}^{z}$ is the Overhauser field operator and $P_{\pm}=\langle{\pm}|S^{z}|\pm\rangle$. The CTs are characterised by $P_{+} \simeq {P}_{-}$ such that central spin decoherence at the \ket{+}$\leftrightarrow$\ket{-} transition is strongly suppressed due to the nearly identical evolutions of the nuclear environment conditioned on the central spin state (i.e.\  $H^{(+)}\simeq H^{(-)}$). Consequently, the back-action of the central spin on the environment~\cite{NV-1,NV-2} is quite small near the CTs (${|P_{+}-P_{-}|}\ll|{P_{+}+P_{-}}|$),
so we may infer that a semiclassical model for quantum decoherence should well reproduce the results from the quantum model.

\begin{figure}
\centering
\includegraphics[width=8.6cm]{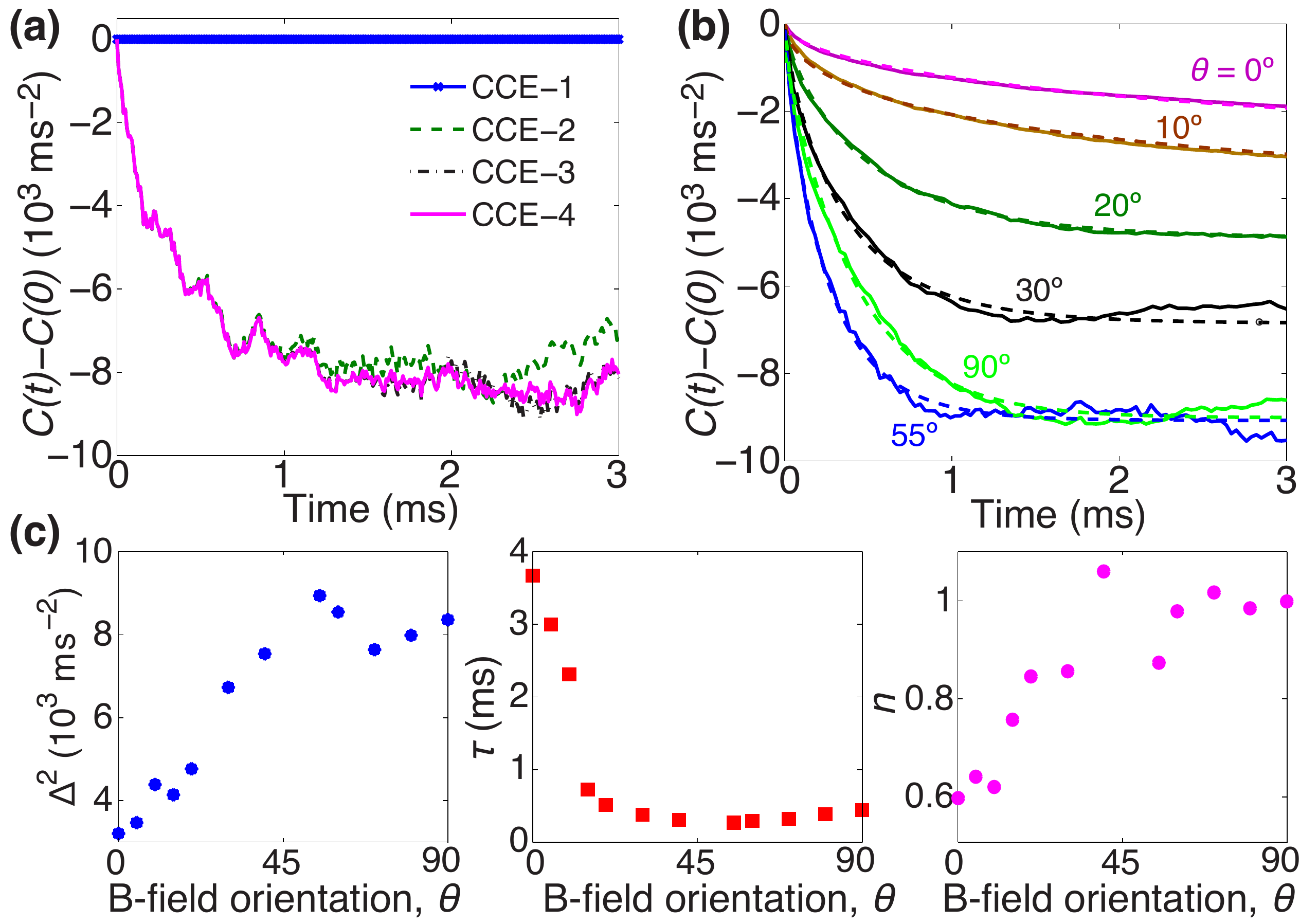}
\caption{(Color online) (a) Relative two-point correlation functions $C(t)-C(0)$ at the CT ($B_{\text{CT}}=79.9$ mT) calculated by different orders of CCE. Here we choose a specific nuclear spin configuration with $\mathbf{B}\parallel[110]$. (b) $C(t)-C(0)$ (solid lines) at the CT for several magnetic field orientations in the $[001]-[110]$ plane with $\theta=0^{\circ}$ corresponding to [001]. Results are obtained by averaging over 50 different nuclear spin configurations. Dashed lines are fits of the form $\Delta^{2}\{\text{exp}[-(|t|/\tau)^{n}]-1\}$. (c) The fitting parameters as functions of $\theta$.}
\label{Ct}
\end{figure}

The two-point correlation function of the nuclear spin noise is defined as
\begin{eqnarray}\label{}
 C(t)=\langle\beta^{z}(t)\beta^{z}(0)\rangle=\langle\text{e}^{iH_{e}t}\beta^{z}(0)\text{e}^{-iH_{e}t}\beta^{z}(0)\rangle,
 \label{}
\end{eqnarray}
where $\langle\cdots\rangle=\text{Tr}[\rho_{b}\cdots]$ denotes the ensemble average over the density matrix $\rho_{b}=\mathbb{I}/2^{M}$ for $M$ nuclear spins at infinite temperature, and $H_{e}=\frac{|P_{+}|+|P_{-}|}{2}\beta^{z}+H_{\text{bath}}$ is the effective Hamiltonian for the nuclear spin bath.
We choose the factor $\frac{|P_{+}|+|P_{-}|}{2}$ instead of $\frac{P_{+}+P_{-}}{2}$ in $H_{e}$ for the following considerations: (i) near the CTs, $P_{+}\simeq{P}_{-}$, so these two factors differ only in their sign which has no effect on the nuclear spin dynamics; (ii) far away from the CTs, $P_{+}\approx-P_{-}\approx{1/2}$, so the chosen factor can better represent the back-action of the central spin on the nuclear spin bath.

In dynamical decoupling (DD) control, a sequence of $N$ $\pi$-flips are applied to the central spin at times $\{t_{1},t_{2}\cdots{t_{N}}\}$ to suppress magnetic noise \cite{DD-1,DD-2}. In this paper we consider  $N$-pulse Carr$-$Purcell$-$Meiboom$-$Gill (CPMG-$N$) control \cite{CPMG-1,CPMG-2} with $t_{k}=(2k-1)/2N$ ($k=1,2,\cdots,{N}$). With the Gaussian approximation \cite{Gauss}, the central spin decoherence under DD control is
\begin{eqnarray}\label{L(t)}
 &&  L(t)=\text{exp}\left[{-\frac{P_{e}^{2}}{2}\int_{0}^{t}\int_{0}^{t}dt_{1}dt_{2}C(t_{1}-t_{2})f(t_{1})f(t_{2})}\right]  \nonumber \\
 && ~~~~~~ =\text{exp}\left[{-\frac{P_{e}^{2}}{2}\int_{-\infty}^{\infty}\frac{d\omega}{2\pi}C(\omega)\frac{F(\omega{t})}{\omega^{2}}}\right],
\end{eqnarray}
where $P_{e}=|P_{+}-P_{-}|$ is the scaling factor, and $C(\omega)=\int_{-\infty}^{\infty}\text{\textit{e}}^{i\omega{t}}C(t)dt$ is the noise spectrum of the nuclear spin noise [since $C(t)=C(-t)$, so $C(\omega)=C(-\omega)$], $f(t)=(-1)^{k}$ for $[t_{k},t_{k+1}]$ is the modulation function, $F(\omega{t})=|\sum_{k=0}^{N}(-1)^{k}(\text{\textit{e}}^{i\omega{t}_{k+1}}-\text{\textit{e}}^{i\omega{t}_{k}})|^{2}$ is the filter function.

\begin{figure}
\centering
\includegraphics[width=8.6cm]{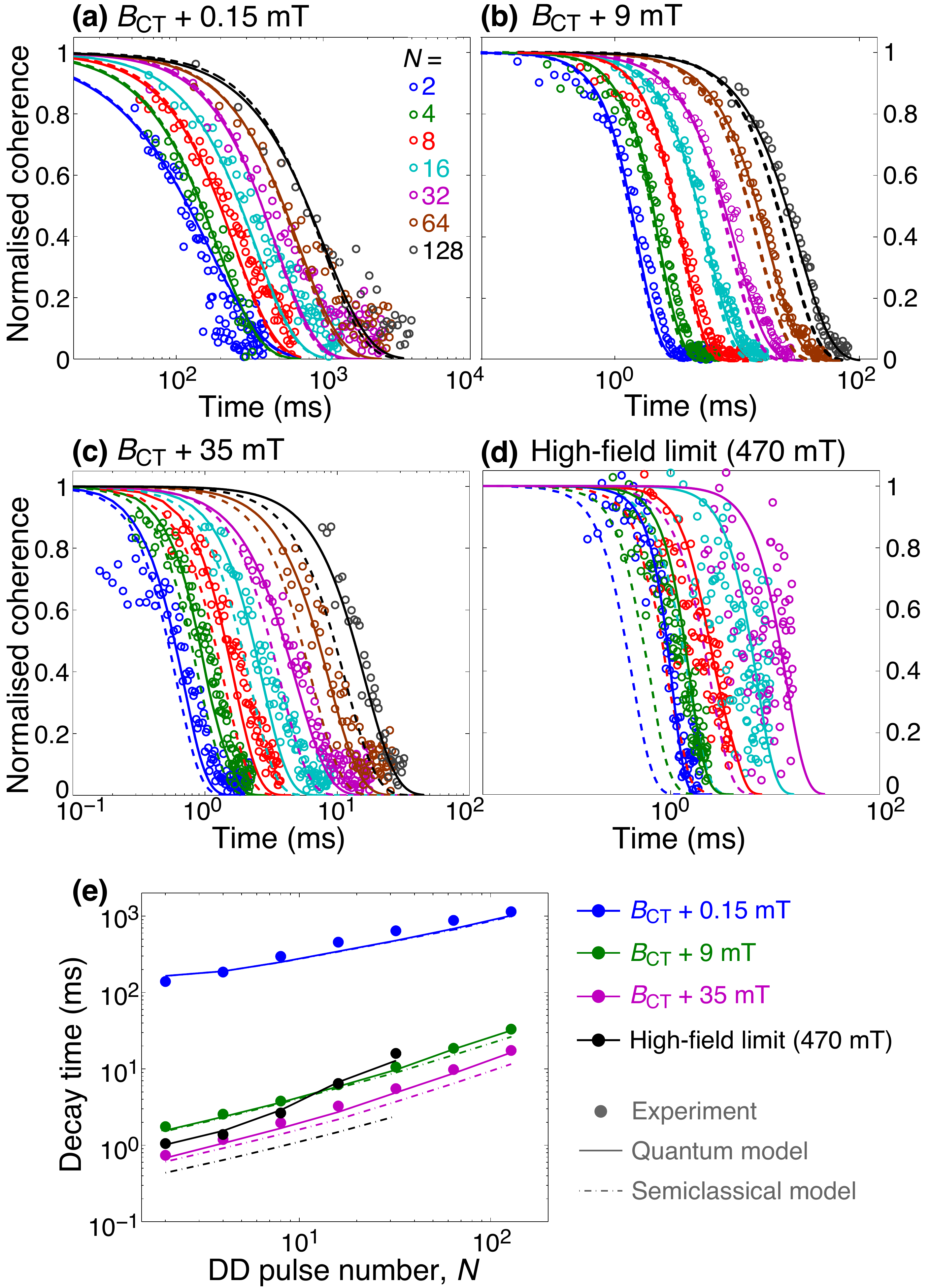}
\caption{ (Color online) Comparisons of electron spin decoherence obtained by the quantum model (solid lines), the semiclassical model (dashed lines)] and the experimental measurement (circles). Measurements were made using (a-c) the \ket{5,-1}$\leftrightarrow$\ket{4,-2} transition at various fields shifted from the CT at 79.9~mT, or (d) the \ket{5,-4}$\leftrightarrow$\ket{4,-5} transition close to the high-field limit at $468.65$~mT.
(e) The spin decoherence time for $T_{\rm 2e}$ under CPMG control obtained from the models and experiment for various fields. Here $N$=2, 4, 8, 16, 32, 64, 128 corresponds to the DD control CPMG-2, XY-4, XY-8, XY-16, (XY-16)$\times$2, (XY-16)$\times$4, (XY-16)$\times$8. In theoretical calculations, CPMG-$N$ is equivalent to XY-DD. The experimental data is corrected to remove effects of instantaneous diffusion and spin relaxation. The theoretical results are obtained by averaging over 100  nuclear spin configurations.}
\label{compare}
\end{figure}

\textit{Characterization of the nuclear spin noise}. The correlation functions $C(t)$ for the natural $^{29}$Si nuclear spin bath in silicon are calculated by the CCE method \cite{quan-CCE,Corre,SI}. As a typical example, we choose to study the central spin transition between $|+\rangle\equiv|5,-1\rangle$ and $|-\rangle\equiv|4,-2\rangle$, using the basis \ket{F,m_F} where $F (=I+S)$ is the total spin and $m_F (= m_S+m_I)$ is its projection. This transition is a CT at $B_{}=79.9$~mT (where $P_{+}=P_{-}=0.0525$).

In Fig.~\ref{Ct}(a), we show the relative correlation function $C(t)-C(0)$ corresponding to the quantum fluctuations of nuclear spin noise calculated for a random nuclear spin configuration with $\mathbf{B}\parallel[110]$.
Since $C(t)$ depends on the specific nuclear spin configuration, we show in Fig.~\ref{Ct}(b) the relative correlation functions averaged over many bath configurations. The results can be well fitted by a stretched exponential decay:
\begin{eqnarray}\label{}
 C(t)=C(0)+\Delta^{2}\{\text{exp}{[-(|t|/\tau)^{n}]}-1\},
 \label{}
\end{eqnarray}
where $\Delta$ is the correlation amplitude, $\tau$ is the correlation time and $n$ is the stretch factor.
In Fig.~\ref{Ct}(c), we show the fitting parameters as functions of the magnetic field orientation. As the field direction varies from
 $[001]\rightarrow[111]\rightarrow[110]$, the correlation amplitude $\Delta$ first increases with $\theta$ (the angle from [001]), reaches a maximum at about $55^{\circ}$ ($[111]$) and then slightly decreases. The stretch factor $n$ has about the same trend as $\Delta$ apart from some oscillations due to systematic fitting errors, while the correlation time $\tau$ has the opposite trend. The dependence of the correlation functions on the magnetic field orientation is due to the anisotropy of the dipolar interaction between nuclear spins and the silicon lattice structure \cite{angu-1,angu-2}, and can be well understood from a microscopic analysis of the nuclear spin dynamics.

 As the major contribution [see \ref{Ct}(a)], the pairwise flip-flop processes in the nuclear spin bath lead to a correlation function as
 \begin{eqnarray}\label{}
  C(t)=C(0)+\sum_{\{i,j\}}\left\{\frac{2Z_{ij}^{2}D_{ij}^{2}}{Z_{ij}^{2}+D_{ij}^{2}}\left[\text{cos}(\omega_{ij}t)-1\right]\right\},
 \label{}
\end{eqnarray}
 where $\omega_{ij}=2\sqrt{Z_{ij}^{2}+D_{ij}^{2}}$ is the noise frequency, $Z_{ij}=\frac{|P_{+}|+|P_{-}|}{4}(A_{i}-A_{j})$ is the energy cost of flip-flop processes, $D_{ij}=\gamma_{n}^{2}(3\text{cos}^{2}\vartheta_{ij}-1)/4|\mathbf{R}_{ij}|^{3}$ is the dipolar interaction strength, $\mathbf{R}_{ij}$ is the displacement between the $i$th and $j$th nuclear spins, and $\vartheta_{ij}$ is the angle between $\mathbf{R}_{ij}$ and $\mathbf{B}$. For a given $Z_{ij}$, the correlation amplitude and noise frequency increase with the dipolar interaction strength $D_{ij}$. When $\mathbf{B}\parallel[001]$, the nearest-neighbor nuclear spin pairs have zero dipolar interaction, so the flip-flop processes mainly occur between the 2rd and 3rd-nearest neighbors. This produces a minimum in $\Delta$ and $n$ and a maximum in $\tau$, as a result of the relatively slow nuclear spin dynamics. When $\mathbf{B}\parallel[111]$, the nearest-neighbor nuclear spin pairs have the strongest dipolar interaction, so $\Delta$ reaches a maximum, $\tau$ a minimum, and $n\approx1$ since the relatively fast nuclear spin dynamics make the noise like a classical Lorentzian noise.

\begin{figure}
\centering
\includegraphics[width=8.6cm]{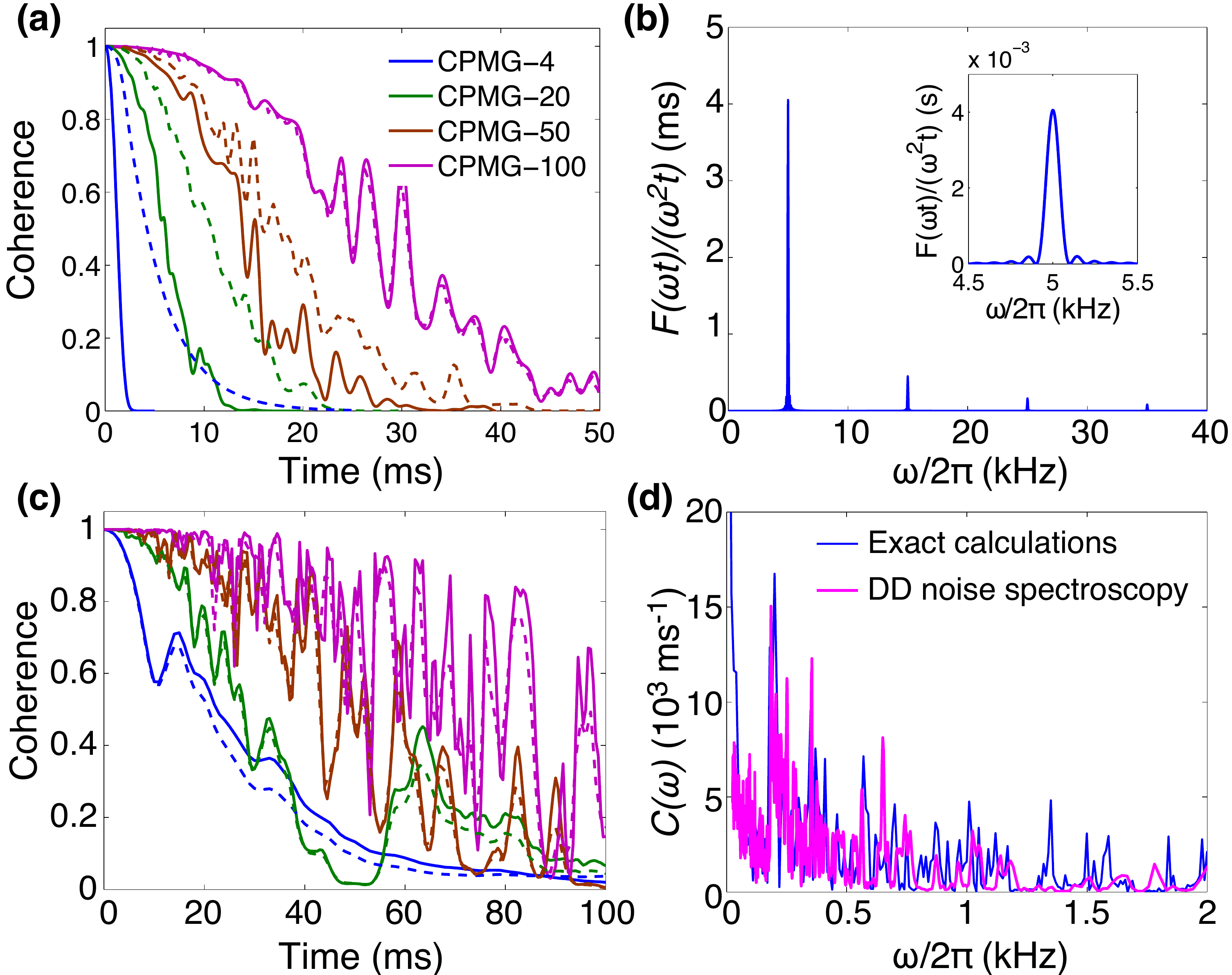}
\caption{(Color online) Calculated electron spin decoherence under an exact quantum model (solid lines) and semiclassical model obtained from noise spectroscopy of the CPMG-100 DD (dashed lines), evaluated (a) far from the CT ($B_{\text{CT}}+1000$ G) and
(c) close to the CT ($B_{\text{CT}}+10$ G).
(b) Filter function $F(\omega{t})/(\omega^{2}t)$ for CPMG-100 noise spectroscopy, with $t=10$~ms.
(d) Comparison of the noise spectrum from the CPMG-100 spectral decomposition in (c) to the exact one from CCE calculations. Here we choose a specific nuclear spin configuration with $\mathbf{B}\parallel[110]$. }
\label{spectral}
\end{figure}

\textit{Comparisons with experiments and quantum model}. To explore the validity of the semiclassical model, we compare in Fig.~\ref{compare} the decoherence obtained from Eq.~(\ref{L(t)}) with both the results from the CCE method \cite{quan-CCE} and the experimental measurements. The measurements were conducted on a $^{\text{nat}}$Si sample doped with $^{209}$Bi at a concentration of 3$\times$10$^{15}$ cm$^{-3}$. The experiments were realized at 4.8 K, which gives an electron spin relaxation time $T_{1e}$ of 3.5 s. The magnetic field was aligned close to the $[\overline{2}41]$ crystal axis. The DD $\pi$ pulses were applied through adiabatic fast passages (hyperbolic secant functions, 20~$\mu s$ in duration spanning 6~MHz close to the CT and 12~MHz in the high-field limit) in order to achieve high-fidelity operations despite the ESR linewidth, which ranged from 6 to 12~MHz at the fields studied here \cite{SI}. Owing to the high-fidelity DD control via the adiabatic passage method, remarkably, we managed to apply more than 100 control pulses and therefore extend the electron spin coherence times to one second in the natural silicon sample. We expect this can be further increased by using optimal magnetic field orientations and higher-order DD.

The results from the quantum model and the experimental data are in good agreement for all the the magnetic fields. However, this is not the case for the semiclassical model. Close to the CT [$B_{\text{CT}}+0.15$ mT in Fig. \ref{compare}(a), where $P_{+}=0.0527$, $P_{-}=0.0521$], the semiclassical coherence, quantum coherence and experimental results coincide for various DD control sequences, indicating that the nuclear spin bath is well described by a classical noise. As the magnetic field is shifted away from the CT [$B_{\text{CT}}+9$ mT in Fig. \ref{compare}(b), where $P_{+}=0.0695$, $P_{-}=0.0340$] the semiclassical model begins to show small deviations from the quantum model for DD with large numbers of pulses. Further still  [$B_{\text{CT}}+35$ mT in Fig. \ref{compare}(c), where $P_{+}=0.1172$, $P_{-}=-0.0198$], there are clear deviations between the semiclassical and quantum models for all levels of DD control, but the time scales remain nearly equal. Finally, for the transitions in the high-field limit [$B=470$ mT in Fig. \ref{compare}(d), where $P_{+}=0.4264$, $P_{-}=-0.5$], the semiclassical model ceases to be valid and shows significant differences from the quantum model and the experiment.
These comparisons demonstrate the classical nature of the nuclear spin noise near the CT. This is understandable since the feedback of the central spin on the evolution of the nuclear spin bath ($\sim\frac{|P_{+}-P_{-}|}{2}$) is largely reduced at the CT.

\textit{Limitations of DD noise spectroscopy method}. Previous studies have adopted the DD noise spectroscopy method to characterize the baths \cite{spec-1,spec-2,spec-3}. The main idea is to use a specific DD control sequence (such as CPMG-$N$ with large $N$) with the filter function approximated as a Dirac delta function at $\omega_{0}=\pm\pi{N}/t$ [see Fig.~\ref{spectral}(b)], i.e., $F(\omega{t})/(\omega^{2}t)\approx\pi[\delta(\omega-\omega_{0})+\delta(\omega+\omega_{0})]$, and following Eq.~(\ref{L(t)}) to determine the noise spectra as $S(\pm\omega_{0})=-2\text{ln}[L(t)]/(tP_{e}^{2})$. This method relies on the validity of the semiclassical model.

However, the noise model may be insufficient when the back-action of the central spin on the environment dynamics is significant. For example, in the $^{\text{nat}}$Si:Bi system, the noise model is invalid for transitions far away from CTs. To demonstrate this point, we use the DD noise spectroscopy method to determine the effective noise spectra corresponding to the CPMG-100 case, and then use the derived noise spectra to calculate the spin decoherence under other DD control sequences.
In Fig.~\ref{spectral}(a) we show the comparisons between the exact decoherence model and the semiclassical model using the DD noise spectroscopy method, and find increasing discrepancies as the pulse number of CPMG-$N$ deviates from 100. In contrast, close to the CTs, the DD noise spectroscopy method can not only reproduce the spin decoherence curves for other DD control [see Fig.~\ref{spectral}(c)], but also well reproduce the exact noise spectrum obtained from CCE calculations [see Fig.~\ref{spectral}(d)]. Here the decoherence profile shows increasingly violent oscillations (corresponding to the relatively high-frequency noise) for a single nuclear spin configuration, reducing the efficiency of DD control at long-time scales. The reason is that when the DD modulation frequency matches the noise frequency, the noise can be amplified rather than suppressed~\cite{mag-2}.

\textit{Summary}. We have presented a semiclassical model to study the decoherence of electron spin qubits in natural silicon near the CTs in $^{\text{nat}}$Si:Bi system. The comparisons of the semiclassical results against the exact quantum results and experimental measurements demonstrate that the nuclear spin bath acts more and more like a classical Gaussian noise as the CTs are approached. Our findings deepen the understanding of spin baths near CTs and are useful for optimizing the DD control in silicon-based quantum computation --- indeed we have already shown here that using DD at CTs, electron spin coherence times of about one second can be measured in natural silicon.

\begin{acknowledgments}
 This work was supported by National Basic Research Program of China (973 Program) under Grant No.\ G2009CB929300, National Natural Science Foundation of China under Grant No.\ 61121491, and Hong Kong RGC/CRF CUHK4/CRF/12G. Work at UCL was supported by the European Research Council under the European Community's
Seventh Framework Programme (FP7/2007-2013)/ERC (Grant No.\ 279781), and by the Engineering and Physical Sciences Research Council (EPSRC) grants EP/K025945/1 and EP/I035536/2. J.J.L.M. is supported by the Royal Society.
\end{acknowledgments}

\end{document}